\documentclass[12pt]{article}

\usepackage[margin=1.0in]{geometry}
\usepackage{parskip}
\usepackage{pstricks,pst-node}
\usepackage{graphicx}

\begin{document}

\title{A Many-Antenna High-Rate Wireless System}
\author{Phil Gossett, Jeremy Thorpe \\
  Bob Nuckolls, Brett Coon, Dan McCloskey, David Chang \\
  Greg Steuck, Paul Rodman, Sasha Levitskiy, Yuan Yuan \\
  \small{cphilgossett@gmail.com} \\
  \small{\{jeremyt,bnuckolls,bwc,dmccloskey,davidchang,gnezdo,rodman,sanek,yuan\}@google.com} \\
  \emph{Google, Inc.}}
\maketitle

\section{Abstract}

We describe a TDD MIMO wireless system designed to operate at high bandwidth and low SNR.  Signals are transmitted as a direct sequence.  In the uplink (Multiple Access Channel), signal detection is done by a space-time whitening filter followed by a matched filter.  In the downlink (Broadcast Channel), precoding is done by the transpose of these filters.

We further describe an implementation of this system that uses an array of 32 antennas to communicate with 32 single-antenna clients simultaneously on the same frequencies between 512-608 and 614-698 MHz.  At close range, all 32 links achieve the full PHY data rate, both uplink and downlink, with less than 1\% Block Error Rate on each link.  The total system rate is 3.8 Gb/s.  The system spectral efficiency is 21.7 b/s/Hz for both uplink and downlink.

We close with some projections to the not-to-distant future.

\section{Introduction}

Claude Shannon showed in 1949 \cite{Shannon49} that capacity in the Gaussian noise channel grows only logarithmically with SNR:

$C = B * log2(1 + P/N)$

Paul Baran pointed out in 1995 \cite{Baran95} that increasing power increases interference.

Yet many recent communication systems, such as 802.11n \cite{WiFi} and LTE \cite{LTE}, try to maximize data rate on a single link by using high order modulation, such as 256-QAM.  This increases the required SNR.  Moreover, the transmitter array typically does not have knowledge of the channel, further increasing the required transmitted power.

By contrast, our system requires a lower SNR at each receiver by using low order modulation, a direct binary sequence.  Moreover, in downlink the basestation uses knowledge of the channel obtained in the uplink to precode, constructing a coherent signal at the intended receiver and maximizing its received power for a given transmitted power allocation.

Our system natively uses wide bandwidth.  This makes it possible to achieve relatively high link capacity, without high order modulation.  It takes practical advantage of having a large number of antennas in the basestation, while requiring clients to have only a single antenna.  Because clients and basestation transmit with low EIRP, it causes minimal interference and is suitable for unlicensed use.  Being able to receive at low SNR makes it less susceptible to interference.

These same properties also make our system less susceptible to and less responsible for self-interference.  This in turn allows a much larger number of antennas than would otherwise be practical, further increasing system efficiency.  By sacrificing capacity on individual links, we can achieve much higher total system capacity.

Our implementation uses the spectrum between 512-608 and 614-698 MHz, with a 32 antenna basestation, and 32 single-antenna clients.  It achieves a total system performance of 1.9 Gb/s uplink plus 1.9 Gb/s downlink, with a system spectral efficiency of 21.7 b/s/Hz for both uplink and downlink.

Quoting from Baran \cite{Baran95}, ``A link is not a system.''  By thinking of wireless communications as a system rather than as individual links, we can achieve much greater efficiencies in the use of the available spectrum.

\section{MIMO Algorithm}

Our system consists of a $M$-antenna basestation capable of communicating simultaneously with $N$ clients, using Time-Division Duplexing.  In the uplink, clients all transmit to the basestation, and in the downlink, the basestation transmits to all the clients.  Downlink precoding uses knowledge of the channel gained in the uplink.

Although the number of antennas and number of clients are independent, in practice it is convenient to let $N = M$.  Such a configuration is nearly optimal for peak data rate with a constrained amount of hardware.  Greater power efficiency can be achieved though by increasing just the number of antennas.

\subsection{MMSE Detection}

\begin{figure}[ht!]
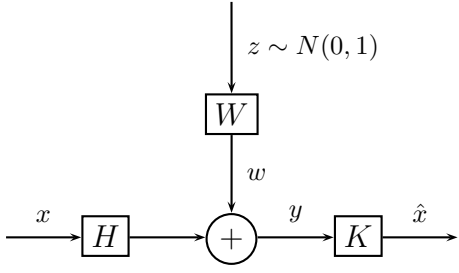

  \begin{psmatrix}[colsep=1cm, rowsep=1cm]
    \pnode{a} & & \pnode{z} & & \\
    \pnode{b} & & \psframebox{$W$} & & \\
    \pnode{c} & \psframebox{$H$} & \pscirclebox{+} & \psframebox{$K$} &
    \ncline{->}{1,3}{2,3}
    \Aput{\footnotesize{$z \sim N(0,1)$}}
    \ncline{->}{2,3}{3,3}
    \Aput{\footnotesize{$w$}}
    \ncline{->}{3,1}{3,2}
    \Aput{\footnotesize{$x$}}
    \ncline{->}{3,2}{3,3}
    \ncline{->}{3,3}{3,4}
    \Aput{\footnotesize{$y$}}
    \ncline{->}{3,4}{3,5}
    \Aput{\footnotesize{$\hat{x}$}}
  \end{psmatrix}
  \caption{MMSE Detection}
  \label{MMSE_Detection}
\end{figure}

Figure~\ref{MMSE_Detection} shows the Minimum Mean Squared Error detection algorithm, where

$K = \frac{H^*}{|H|^2+E(WW^*)} = \frac{H^*}{E(YY^*)}$ .

K is a Wiener filter \cite{Wiener49}.  Here, H is the transfer function of the signal channel, $E(WW^*)$ is the power spectrum of the noise and $E(YY^*)$ is the power spectrum of the signal y.

The transfer function is estimated by integrating and windowing the correlation between the received signal and pilot signal (a long pseudo-random sequence).  The spectrum $E(YY^*)$ is estimated in the frequency domain and is averaged both over time and over a range of adjacent frequency bins.

\subsection{Whitening and Matched Filters}

\begin{figure}[ht!]
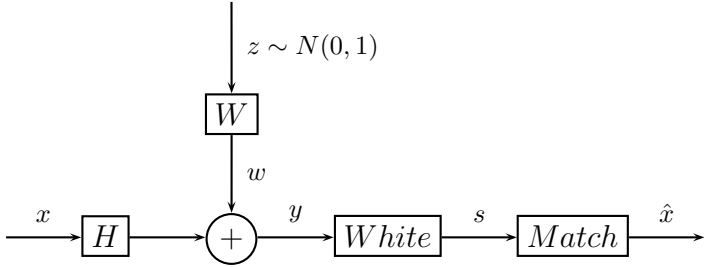

  \begin{psmatrix}[colsep=1cm, rowsep=1cm]
    \pnode{a} & & \pnode{z} & & & \\
    \pnode{b} & & \psframebox{$W$} & & & \\
    \pnode{c} & \psframebox{$H$} & \pscirclebox{+} & \psframebox{${White}$} & \psframebox{${Match}$} &
    \ncline{->}{1,3}{2,3}
    \Aput{\footnotesize{$z \sim N(0,1)$}}
    \ncline{->}{2,3}{3,3}
    \Aput{\footnotesize{$w$}}
    \ncline{->}{3,1}{3,2}
    \Aput{\footnotesize{$x$}}
    \ncline{->}{3,2}{3,3}
    \ncline{->}{3,3}{3,4}
    \Aput{\footnotesize{$y$}}
    \ncline{->}{3,4}{3,5}
    \Aput{\footnotesize{$s$}}
    \ncline{->}{3,5}{3,6}
    \Aput{\footnotesize{$\hat{x}$}}
  \end{psmatrix}
  \caption{Whitening and Matched Filters}
  \label{Whitening_and_Matched_Filters}
\end{figure}

Figure~\ref{Whitening_and_Matched_Filters} shows the SISO (and client) detection algorithm, where

${White} = \frac{1}{\sqrt{|H|^2 + E(WW^*)}} = \frac{1}{\sqrt{E(YY^*)}}$ and

${Match} = \frac{H^*}{\sqrt{|H|^2 + E(WW^*)}} = \frac{H^*}{\sqrt{E(YY^*)}}$

The K of Figure~\ref{MMSE_Detection} is split into whitening and matched filters.  The whitening filter normalizes the power spectrum, so that the $E(SS^*)$ has uniform spectral density.  The matched filter is the complex conjugate of the transfer function of the channel including the whitening filter.

\subsection{Space-Time Whitening}

We extend this to the MIMO uplink (Multiple Access Channel \cite{Goldsmith03}) case with space-time whitening.

The diagram and equations for the MIMO case look identical to the SISO case (Figure~\ref{Whitening_and_Matched_Filters}), with terms such as $x$, $y$, $s$, interpreted as vectors, and $YY^*$, $WW^*$, and $H$ as matrices.

Here, $Y$ is a vector of frequency domain representations of the received signal vector $y$.  The expectation of $YY*$ is the covariance matrix of $Y$, which again is integrated and binned in the frequency domain.

The whitening filter is the matrix inverse of the square root of the covariance matrix of $Y$.  The matched filter is the conjugate transpose of the channel matrix $H$, again including the whitening filter.

The covariance matrix is Hermitian positive semi-definite (by construction).  So the square root can be implemented as the Cholesky decomposition, which results in the whitening filter being a triangular matrix of transfer functions.  This triangular form is numerically stable \cite{Turing48} and convenient for implementation.

\subsection{Linear Precoding}

\begin{figure}[ht!]
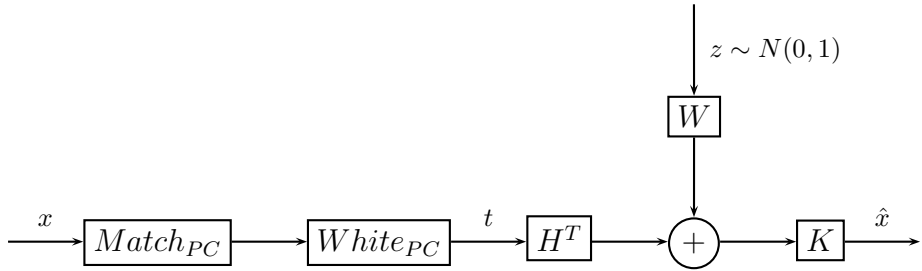

  \begin{psmatrix}[colsep=1cm, rowsep=1cm]
    \pnode{a} & & & & \pnode{z} & & \\
    \pnode{b} & & & & \psframebox{$W$} & & \\
    \pnode{c} & \psframebox{${Match}_{PC}$} & \psframebox{${White}_{PC}$} & \psframebox{$H^T$} & \pscirclebox{+} & \psframebox{$K$} &
    \ncline{->}{1,5}{2,5}
    \Aput{\footnotesize{$z \sim N(0,1)$}}
    \ncline{->}{2,5}{3,5}
    \ncline{->}{3,1}{3,2}
    \Aput{\footnotesize{$x$}}
    \ncline{->}{3,2}{3,3}
    \ncline{->}{3,3}{3,4}
    \Aput{\footnotesize{$t$}}
    \ncline{->}{3,4}{3,5}
    \ncline{->}{3,5}{3,6}
    \ncline{->}{3,6}{3,7}
    \Aput{\footnotesize{$\hat{x}$}}
  \end{psmatrix}
  \caption{Linear Precoding}
  \label{Linear_Precoding}
\end{figure}

Figure~\ref{Linear_Precoding} shows the linear precoding (Broadcast Channel \cite{Goldsmith03}) algorithm, where

${Match}_{PC} = \left( \frac{1}{\sqrt{E(YY^*)}} \right) ^T = {Match}^T$ and

${White}_{PC} = \left( \frac{H^*}{\sqrt{E(YY^*)}} \right) ^T = {White}^T$ .

The channel matrix in the downlink direction is the transpose of the channel matrix in the uplink direction.  So the downlink precode transfer function is the transpose of the uplink detection transfer function ${White}$ composed with ${Match}$ (from Figure~\ref{Whitening_and_Matched_Filters}), which is ${Match}_{PC}$ composed with ${White}_{PC}$.  $K$ is simply the whitening and matched filters for each of the client nodes as in the SISO case.

\section{Implementation}

\begin{figure}[ht!]
  \begin{psmatrix}[colsep=0.5cm,rowsep=0.5cm]
    & & & \psframebox{$pilot$} \\
    \pnode{b} & \psframebox{$encode$} & \psframebox{$\Pi$} & \pscirclebox{+} & \psframebox{$precode^N_M$} & \psframebox{$mask$} & \psframebox{$DAC$} & \psframebox{$mod$} & \pspolygon[](0,0)(0,1)(-.25,1.25)(.25,1.25)(0,1) \\
    & & & & & & & & & \psframebox[linestyle=dashed]{$H^M_N$} \\
    \pnode{d} & \psframebox{$decode$} & \psframebox{$\Pi^{-1}$} & & \psframebox{$detect^N_M$} & & \psframebox{$ADC$} & \psframebox{$demod$} & \pspolygon[](0,0)(0,1)(-.25,1.25)(.25,1.25)(0,1) \\
    & & & & & & $N(0,1)$ & \psframebox[linestyle=dashed]{$W^N_N$}
    \ncline[linestyle=dashed]{->}{2,9}{3,10}
    \ncline[linestyle=dashed]{->}{3,10}{4,9}
    \ncline[linestyle=dashed]{->}{5,8}{4,9}
    \ncline[linestyle=dashed]{->}{5,7}{5,8}
    \naput{\footnotesize{$z_N$}}
    \ncline{->}{1,4}{2,4}
    \naput{\footnotesize{$p_N$}}
    \ncline{->}{2,1}{2,2}
    \naput{\footnotesize{$m_N$}}
    \ncline{->}{2,2}{2,3}
    \naput{\footnotesize{$d_N$}}
    \ncline{->}{2,3}{2,4}
    \naput{\footnotesize{$x_N$}}
    \ncline{->}{2,4}{2,5}
    \ncline{->}{2,5}{2,6}
    \naput{\footnotesize{$y_M$}}
    \ncline{->}{2,6}{2,7}
    \ncline{->}{2,7}{2,8}
    \ncline{->}{2,8}{2,9}
    \naput{\footnotesize{$t_M$}}
    \ncline{->}{4,9}{4,8}
    \nbput{\footnotesize{$r_N$}}
    \ncline{->}{4,8}{4,7}
    \ncline{->}{4,7}{4,5}
    \nbput{\footnotesize{$\widehat{y_N}$}}
    \ncline{->}{4,5}{4,3}
    \nbput{\footnotesize{$\widehat{x_M}$}}
    \ncline{->}{4,3}{4,2}
    \nbput{\footnotesize{$\widehat{d_M}$}}
    \ncline{->}{4,2}{4,1}
    \nbput{\footnotesize{$\widehat{m_M}$}}
  \end{psmatrix}
  \par\vspace{\baselineskip}
  \begin{tabular}{|l|l|}
    \hline
    {$encode$}      & FEC encoder                         \\
    {$\Pi$}         & Permutation                         \\
    {$pilot$}       & Pilot lookup table                  \\
    {$precode^N_M$} & Precoder                            \\
    {$mask$}        & Transmit mask filter                \\
    {$DAC$}         & Digital to analog converter         \\
    {$mod$}         & Synchronous heterodyne modulator    \\
    {$H^M_N$}       & Signal channel matrix (M Tx, N Rx)  \\
    {$W^N_N$}       & Noise channel matrix (N Rx)         \\
    {$demod$}       & Synchronous heterodyne demodulator  \\
    {$ADC$}         & Analog to digital converter         \\
    {$detect^N_M$}  & Detector                            \\
    {$\Pi^{-1}$}    & Inverse permutation                 \\
    {$decode$}      & FEC decoder                         \\
    \hline
  \end{tabular}
  \caption{Block Diagram}
  \label{Block_Diagram}
\end{figure}

Figure~\ref{Block_Diagram} shows a block diagram of the system.  In the uplink direction, the precoder is an identity (i.e., the clients don't precode).  In the downlink direction, the detector is diagonal (i.e., the clients detect independently).  Note that the signal channel matrix $H^M_N$ is transposed between uplink and downlink, and that the noise channel matrix $W^N_N$ is different in the two directions.  The latter is why non-trivial detection is still required in the downlink.

\subsection{Analog/Digital Conversion}

For both the multiple-antenna array and the single-antenna clients, the analog (per antenna) is essentially the same.

At the transmitter, the processed signal is first Digital-to-Analog Converted, then modulated, then passed through a Power Amplifier.  The modulation is a synchronous heterodyne, shifting the baseband frequency to the desired RF frequency band.

At the receiver, the signal is first passed through a Low Noise Amplifier, then demodulated, then Analog-to-Digital Converted for further processing.  The demodulation is again a synchronous heterodyne, shifting from the RF frequency band back to baseband.

The digital signals are processed as Tukey-windowed blocks of 8K samples each, at a Sample Rate of 488.8 MHz.  The synchronous Local Oscillator is the same 488.8 MHz sample clock, moving baseband to 488.8-733.2 MHz.  A combination of analog and digital filtering removes all but 512-608 and 614-698 MHz by better than -55 dB, leaving a Suppressed-Carrier Single-SideBand signal.

Our system uses Time-Division Duplexing.  In the uplink, clients all transmit to the basestation, and in the downlink, the basestation transmits to all the clients.  Each TDD frame consists of 31 blocks of uplink, 31 blocks of downlink, and 2 blocks of guard.

\subsection{Uplink}

In the uplink direction, each of the clients encodes a message $m$ using an LDPC \cite{Gallager63} encoder with repetition to produce $d$, and then permutes (to reduce the effects of InterSymbol Interference) to produce $x$.  Each then forms a Direct Sequence signal $y$ that contains a sum (superposition) of client pilot (a long pseudo-random sequence) and coded data.  Each then masks and modulates to produce the transmitted signal $t$.

The basestation first demodulates the incoming signal $r$ on all antennas to produce $y$.  It then space-time whitens to obtain $s$.  It then computes the correlation of the known pilots with each $s$ to obtain an estimated channel matrix $h$.  It applies a matched filter derived from $h$ to $s$ to obtain $\hat{x}$.  (See Figure~\ref{Whitening_and_Matched_Filters}.)  $\hat{x}$ is inverse-permuted to obtain $\hat{d}$ and decoded by accumulation and an LDPC decoder to obtain $\hat{m}$.  The client timing advance is derived from the pilot as seen at the basestation.

\subsection{Downlink}

In the downlink direction, the basestation encodes for each client a message $m$ using an LDPC encoder with repetition, and then permutes (to reduce the effects of InterSymbol Interference) to produce $x$.  It then forms a vector of Direct Sequence signals that contains a sum (superposition) of client pilots (long pseudo-random sequences) and coded data.  These signals are then precoded with the transpose of the space-time whiten and matched filters to produce the signal $y$.  (See Figure~\ref{Linear_Precoding}.)  It then masks and modulates to produce the transmitted signals $t$.

Each of the clients first demodulates the incoming signal $r$ to produce $y$.  It then whitens to obtain $s$.  Each then computes the correlation of the known pilot with $s$ to obtain an estimated channel $h$.  Each applies a matched filter derived from $h$ to $s$ to obtain $\hat{x}$.  (See Figure~\ref{Whitening_and_Matched_Filters}.)  $\hat{x}$ is inverse-permuted to obtain $\hat{d}$ and decoded by accumulation and an LDPC decoder to obtain $\hat{m}$.  The client phase lock is derived from the pilot as seen at the client.

\subsection{Calibration}

Note that for the Linear Precoding to work, the basestation analog has to be calibrated.  This is accomplished by transmitting (round-robin) a pseudo-random sequence on each antenna of the 32 antenna array while receiving on the other 31 antennas, and using correlations of that sequence to estimate the transfer function for each pair of antennas.  The phase differences are then corrected in the transmit mask for each of the 32 transmitters in the array.

\section{Performance}

\begin{figure}[htb]
  \centering
  \includegraphics[width=0.8\textwidth]{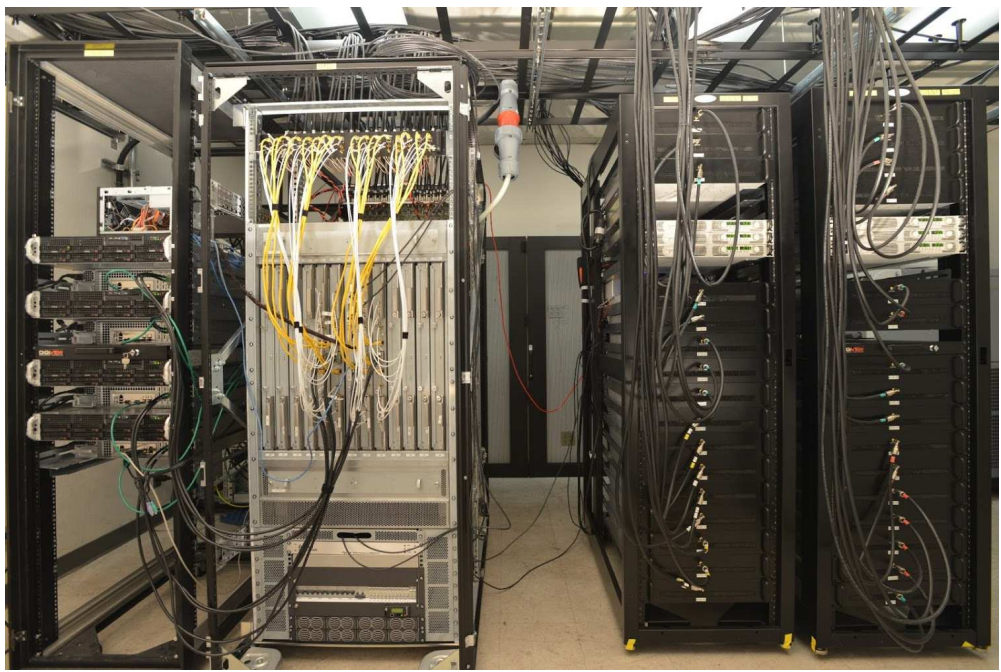}
  \caption{Prototype Base Station and 32 Clients}
  \label{Machine_Room}
\end{figure}

\begin{figure}[htb]
  \centering
  \includegraphics[width=0.8\textwidth]{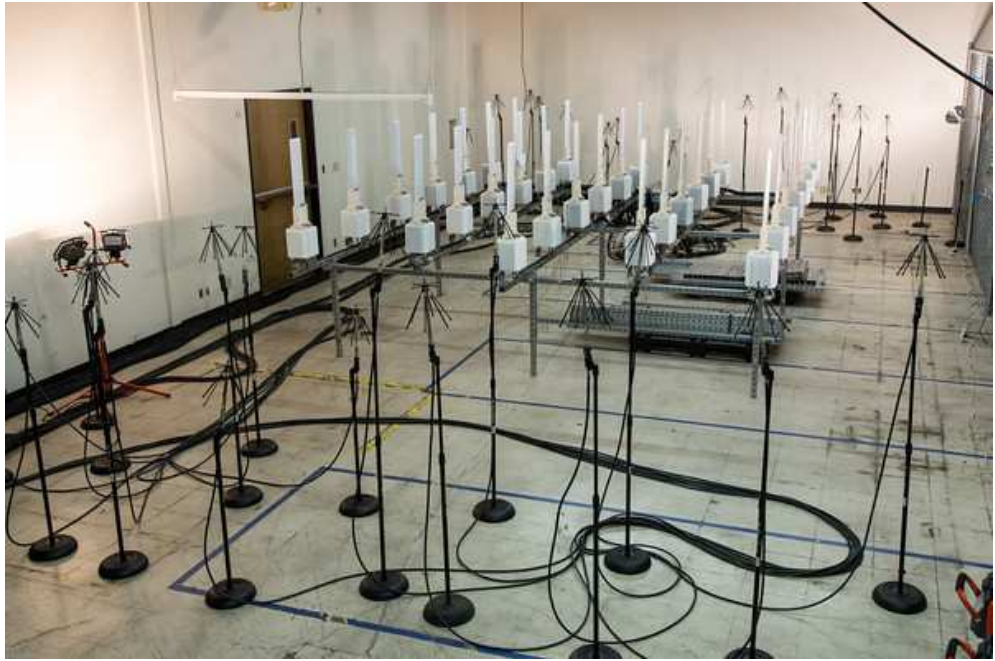}
  \caption{Test Facility}
  \label{Test_Facility}
\end{figure}

Figure~\ref{Machine_Room} shows the prototype hardware for our 32 antenna base station (the leftmost two racks) and 32 clients (16 each in the rightmost two racks).  The prototype base station consists of 337 FPGAs and each of the 32 clients consist of 2 FPGAs.

Figure~\ref{Test_Facility} shows the approximate physical layout of our indoor over-the-air test facility.  The total dimensions are approximately 14 m long by 7 m wide by 5 m high.  The 32 antenna array is in the middle of the facility, with the 32 client antennas distributed on either end.  The average distance between any element of the antenna array and any client antenna is approximately 6 m.  The exact antenna placement is non-critical, down to a minimum spacing of a half-wavelength (30 cm).  No effort was made to make the facility anechoic.

With 100 mW total power on the 32 antenna array, and 1 mW on each of the 32 clients, we have more than enough SNR to achieve our maximum designed data rate, with post-FEC Block Error Rate less than 1\% on each link.  This rate uses a half-rate LDPC, concatenated with 2x repetition.  So our instantaneous data rate at sufficient SNR is 488.8 {Msamples/second} / 4 {samples/bit} = 122.2 {Mb/s}.  With each direction using 31 of 64 blocks per TDD frame, the per link data rate is 59.2 {Mb/s} for each uplink and downlink.  The total system rate is 3.8 Gb/s.

Our in-band signal is between 512-608 and 614-698 MHz, a total of 180 MHz.  So our per link spectral efficiency is 122.2 {Mb/s} / 180 {MHz} = 0.69 b/s/Hz.  Across 32 clients, the system spectral efficiency is 21.7 b/s/Hz.

\section{Discussion}

We have shown that by the combination of low SNR, high order MIMO with precoding, and wide bandwidth, we can achieve significantly greater efficiencies in the practical use of the available spectrum.  Using TDD between 512-608 and 614-698 MHz, with a 32 antenna array, and 32 clients, this combination of properties lets our implementation achieve a total system performance of 1.9 Gb/s uplink plus 1.9 Gb/s downlink.  The system spectral efficiency is 21.7 b/s/Hz for both uplink and downlink.

Significantly increasing spectral efficiency can only come either with greatly higher SNR, which is impractical, or with larger numbers of antennas, which has proved difficult to achieve at high SNR.  Every indication is that increasing the number of antennas at low SNR in a system such as ours is relatively straightforward.

The cost of our system is dominated by the digital computation of and multiplication by $H$ , which at a lower level is dominated by matrix-vector multiplies and divides.  These operations are quadratic in cost with the number of antennas.  (The Cholesky decomposition, which is cubic, is greatly decimated in both time and frequency, and it does not dominate in our implementation.)

The design of our demonstrated 32 antenna system began in 2009, and was limited in performance by then-available FPGAs.  A 64 antenna system is well within reach with FPGAs available at the time that the system was demonstrated in 2012.  A 128 antenna system is achievable with current-generation (22 nm) custom ASICs.

As long as Moore's Law \cite{Moore65} holds (doubling the number of transistors every 2 years), the number of antennas can double every 4 years for the foreseeable future.  Spectral efficiency will double as well.

By the end of the decade, systems with hundreds of antennas and spectral efficiencies in the hundreds of bits per second per Hertz will become practical.

\bibliographystyle{plain}
\bibliography{paper}

\end{document}